\documentclass[conference]{IEEEtran}
\IEEEoverridecommandlockouts

\usepackage{cite}
\usepackage{amsmath,amssymb,amsfonts}
\usepackage{algorithmic}
\usepackage{graphicx}
\usepackage{textcomp}
\usepackage{subcaption}
\usepackage{comment}
\usepackage{xcolor}
\usepackage{siunitx}
\def\BibTeX{{\rm B\kern-.05em{\sc i\kern-.025em b}\kern-.08em
    T\kern-.1667em\lower.7ex\hbox{E}\kern-.125emX}}

\begin{document}

\title{A Self-Calibrating SDR for High Fidelity Beam- and Null-forming Arrays 
\thanks{
This work was partly supported by the Institute of Information \& Communications Technology Planning \& Evaluation(IITP)-Open RAN Education and Training Program grant funded by the Korea government(MSIT) (IITP-2026-RS-2024-00429088); in part, by NSF grants 2345139, 2148293,
2133662, 1952180, and 1904648; the NTIA Public Wireless Innovation
Fund, and the industrial affiliates of NYU WIRELESS.}
}

\author{
\IEEEauthorblockN{Yongjun Kim$^*$, Aditya Dhananjay$^\dagger$$^\ddagger$, Sundeep Rangan$^\ddagger$, Sachin Shetty$^\diamond$, \\C. Nicolas Barati$^\diamond$, Michael Zappe$^\dagger$, Kimberly Gold$^\circ$, and Junil Choi$^*$}
\IEEEauthorblockA{$^*$KAIST School of Electrical Engineering, Daejeon, South Korea\\
$^\dagger$Pi-Radio, Brooklyn, NY, USA\\
$^\ddagger$NYU Tandon School of Engineering, Brooklyn, NY, USA\\
$^\diamond$Old Dominion University Department of Electrical and Computer Engineering, Norfolk, VA, USA\\
$^\circ$NAVSEA Crane Division, Crane, IN, USA \\
Email: \{yongjunkim, junil\}@kaist.ac.kr, 
\{aditya.dhananjay, michael.zappe\}@pi-rad.io \\
srangan@nyu.edu, \{sshetty, cbaratin\}@odu.edu, kimberly.d.gold2.civ@us.navy.mil
}
}

\maketitle

\begin{abstract}
Null forming is increasingly essential in modern wireless systems for spectrum‑sharing,  anti‑jamming, and covert communications in contested and congested environments. Achieving deep nulls, however, is far more demanding than conventional beam steering: nulls are intrinsically narrow, and even small phase, timing, or gain mismatches across RF chains can significantly degrade suppression. This work develops and validates a self‑calibrating SDR architecture tailored for high‑fidelity null forming using a compact reference transmitter directionally coupled to the antenna feeds. We demonstrate the effectiveness of the approach through simulation and experimental measurements on an SDR platform operating from 3.0 to 3.5\, \si{GHz}, a band of growing importance for Department of Defense spectrum‑sharing initiatives.
\end{abstract}


\section{Introduction}
Null forming, where beams are adjusted to create nulls in particular directions, is vital in a range of wireless applications, including covert communications \cite{kong2020simultaneous}, anti-jamming \cite{kumarasiri2025rf}, and spectrum sharing \cite{tashiro2022nullforming, kang2024cellular,jia2025joint}, particularly in non-terrestrial networks and in the mid-band and upper mid-band. A particular practical challenge in null forming is \emph{array calibration}:  Nulls tend to be much more narrow than the main beam. Hence, even small errors in the calibration can result in a significant degradation of the nulling level.
For example, \cite{Madani:2021} showed that nulling performance degrades noticeably once calibration errors exceed about $5^\circ$, and \cite{Wang:2021} reported measured null depths of about $30$~dB for amplitude and phase deviations within $\pm0.5$~dB and $\pm5^\circ$, respectively, even though the mainlobe remains accurate within $\pm5$~dB and $40^\circ$.

Calibration is particularly challenging in software-defined radio (SDR)-based wireless communication systems due to the relatively large hardware impairments. Practical radio-frequency (RF) chains are not perfectly identical, and such mismatches reduce coherent combining and degrade beamforming gain \cite{kim:2019}. Additionally, conventional calibration methods often require costly laboratory instruments such as spectrum analyzers and vector network analyzers (VNA), for instance, two-step calibration through VNA-assisted RF measurements \cite{Wang:2020}, or calibration with separate local oscillators \cite{Dhananjay:2021} provided limited calibration accuracy due to phase synchronization errors, which becomes a major bottleneck in distributed systems \cite{Rogalin:2014}.

In this work, we analyze and experimentally validate a  \emph{self-calibration} in which an SDR estimates and compensates for its own hardware-induced offsets using signals generated and observed by the device itself, i.e., using a loopback structure \cite{Deng:2020}. Self-calibration provides several practical advantages for real deployments. Most importantly, self-calibration does not require expensive external equipment, enhancing the accessibility of SDR-based wireless communication research. Second, over-the-air reference transmissions and separated local oscillator (LO) clocks are no longer required, so unexpected signal distortion due to the wireless channel and clock offset is avoided, increasing the robustness of the calibrated system during repeated operation \cite{Benzin:2017}. Unlike prior self-calibration studies that mainly addressed relative phase/amplitude mismatch \cite{Guo:2020} or I/Q imbalance \cite{De:2008}, our work uses a physically structured timing/phase/gain-offset model that is directly matched to a practical self-calibration SDR and compensated via lightweight digital filters.

The contributions of this paper are summarized as follows:
\begin{itemize}
    \item \emph{Directionally coupled self-calibration board design}:  We develop a simple self-calibration board for the receiver array where a single
    reference transmitter transmits a known signal through a Wilkinson divider that is then directionally coupled to the receiver antenna feeds (see Fig.~\ref{fig:rx self calibration}).  Importantly, the transmission lines from the reference to the coupler can be length-matched so that they are common to all elements. This matching enables the mismatches across different receiver channels to be isolated.
    \item \emph{Array calibration algorithm}: Using the transmit signals, we propose a lightweight self-calibration procedure that estimates the per-channel timing, phase, and gain curve offsets, then constructs a finite impulse response (FIR) compensation filter.
    \item \emph{Experimental validation}:  We validate the effectiveness of the proposed procedure through both simulation data and experimental measurements by comparing the pre- and post- calibrated beampatterns in a null forming scenario. For the experiment, we implement the proposed method on an SDR designed for self-calibration at 3.0 to 3.5\,\si{GHz} -- a band that is a particular focus for the Department of Defense (DoD) spectrum sharing \cite{NTIA2024SpectrumSharing35GHz}.
\end{itemize}

The remainder of the paper is organized as follows. Section~\ref{sec:calibration problem} formulates the self-calibration problem and states its objective. Section~\ref{sec:calibration procedure} presents the FIR compensation filter design and the offsets estimation procedure. Section~\ref{sec:hardware description} describes the SDR architecture designed for self-calibration. Section~\ref{sec:self-calibration validation} evaluates the method through null forming beampatterns. Finally, Section~\ref{sec:conclusion} concludes the paper.

\section{Calibration Problem}
\label{sec:calibration problem}
Consider a fully digital receiver array with $M$ channels. Let $r_m[n]$ denote the complex baseband signal observed at the $m$-th channel at discrete-time index $n$. When the array receives a plane wave from angle $\theta$, the received signal at channel $m$ is modeled as
\begin{align}
    r_m[n] = a_m(\theta)(h_m[n] * u[n]) + w_m[n],
    \label{eq:operational_model}
\end{align}
where $a_m(\theta)$ is the array response corresponding to the $\theta$ at the $m$-th receiver channel, $u[n]$ is the received baseband waveform, $h_m[n]$ is the effective baseband response of the $m$-th receiver channel, $w_m[n]$ is additive noise with noise power $\sigma^2$, and $*$ denotes linear convolution.

The channel $h_m[n]$ captures the hardware impairments of the $m$-th receive channel. While all channels share a common response in an ideal array, in practice, the responses $\{h_m[n]\}_{m=1}^M$ are generally non-identical because nominally identical RF chains still exhibit channel-dependent timing, phase, and gain curve mismatches due to manufacturing tolerances, unequal electrical path lengths in the routing network, and temperature-dependent drift in the RF chain \cite{Delfini:2023}\cite{Vaghefi:2020}. These per-channel mismatches are problematic for multi-antenna SDR because array processing techniques rely on alignment among channels. In particular, coherent combining and null forming are degraded when the channels are mismatched, localization accuracy deteriorates when the relative phase and delay responses are distorted, and reciprocity-based beamforming at a time-division duplexing system is not guaranteed when the hardware responses are not calibrated. Also, recall that $h_m[n]$ has a frequency-selective transfer function because the practical power amplifier (PA)'s gain is not flat across the signal bandwidth \cite{Tkacenko:2010}, resulting in-band distortion. The calibration problem is therefore to estimate $\{h_m[n]\}_{m=1}^M$, which is channel-dependent and frequency selective, and to compensate for them so that all channels align to a common target response.

\begin{figure}[t]
    \centering
    \includegraphics[width=0.85\columnwidth]{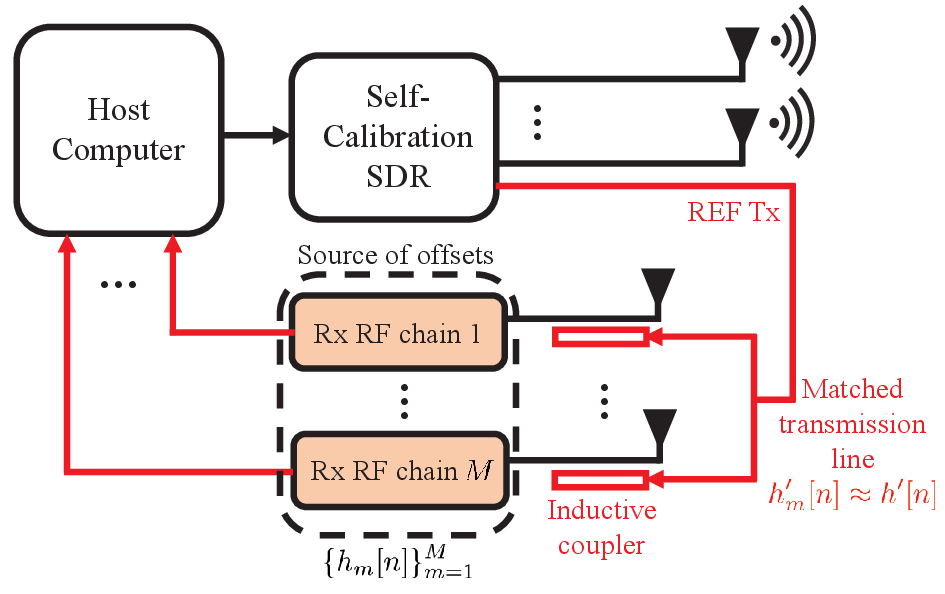}
    \vspace{-0.5em}
    \caption{Receive array self-calibration}
    \label{fig:rx self calibration}
    \vspace{-2em}
\end{figure}

To estimate these responses without external laboratory equipment, we consider a self-calibration mode in which a known reference signal is injected into all receive channels through a common on-board calibration structure, as shown in Fig.~\ref{fig:rx self calibration}. Let $x[n]$ denote the known calibration waveform. During self-calibration, the observation at channel $m$ is modeled as
\begin{align}
    y_m[n] = h_m[n] * h'_m[n] * x[n] + w_m[n],
    \label{eq:self_cal_general_model}
\end{align}
where $h'_m[n]$ denotes the impulse response of the calibration path from the reference source to the $m$-th channel. By careful printed circuit board (PCB) design, the calibration-path responses can be made nearly identical across channels. Hence, we approximate $h'_m[n] \approx h'[n] \text{, }\forall m$. Under this approximation, the dominant channel-dependent variation in the self-calibration measurements is induced by the RF chain responses $\{h_m[n]\}_{m=1}^M$. Accordingly, the calibration problem in this paper is to use the self-calibration measurements $\{y_m[n]\}_{m=1}^M$ from the known calibration waveform $x[n]$ to estimate the offsets observed in $h_m[n]$, and to construct FIR compensation filters $\{f_m[n]\}_{m=1}^M$ that align the channels. Calibration of the transmit array can be considered similarly.

\section{Calibration Procedure}
\label{sec:calibration procedure}

\subsection{Calibration Filter Design}
\label{subsec:calibration filter design}

To estimate the channel response, we first send a pilot QPSK waveform $x[n]$, $n=0,1,\cdots, N-1$. The signal is transmitted repeatedly so that we can estimate the channel frequency response $H_m[k]$ with a discrete Fourier transform (DFT), modeled as
\begin{align}
    H_m[k] = G_m[k]e^{-j\frac{2\pi k}{N}\tau_m}e^{j\phi_m},
    \label{eq:effective gain}
\end{align}
where $k = 0,1,\cdots, N-1$ is a frequency bin index, $\tau_m$ is a per-channel timing offset, $\phi_m$ is a per-channel phase offset, and $G_m[k]$ is a per-channel frequency response at the $k$-th frequency bin. Our objective is to design an FIR calibration filter $f_m[n]$ with frequency response $F_m[k]$ such that $F_m[k]H_m[k] \approx G_0$, where $G_0$ is a common scalar target gain. First, we construct an FIR timing and phase offsets compensator $d_m[n]$ with its frequency response $D_m[k]$. Second, we design an FIR equalizer $q_m[n]$ with frequency response $Q_m[k]$ to flatten the remaining frequency-selective gain. The final FIR filter is therefore $f_m[n] = q_m[n] * d_m[n]$. After obtaining the calibration FIR filter, receive array compensation is implemented in the time domain as $\tilde{y}_m[n] = f_m[n] * y_m[n]$. Although it is possible to design an FIR filter that directly equalizes \eqref{eq:effective gain}, a two-stage approach is adopted since a direct FIR filter that simultaneously learns phase distortion and the residual frequency-selective gain mismatch is inefficient with a limited number of taps.

Suppose that $\hat{\tau}_m$ and $\hat{\phi}_m$ are available. Then, the timing and phase offsets compensator frequency response is $D_m[k] = e^{j\frac{2\pi k}{N}\hat{\tau}_m}e^{-j\hat{\phi}_m}$. Since the post-processing of the received signal is based on linear FIR filtering, we design a standard fractional-delay filter $d_m[n]$ with odd configurable length $L_d$ by shifting the impulse response to the center tap, truncating it, and applying a window function to reduce the Gibbs phenomenon. Next, we design $q_m[n]$ with configurable length $L_q$. Let $\mathbf{q}_m=[q_m[0],\ldots,q_m[L_q-1]]^{\mathsf{T}}$ denote the FIR coefficients, $\mathbf{A}\in\mathbb{C}^{N\times L_q}$ be the partial DFT matrix with entries $[\mathbf{A}]_{k,n}=e^{-j\frac{2\pi k}{N}n}$, $\mathbf{g}_0 = G_0\mathbf{1}_N$, and $\hat{\mathbf{G}}_m=[\hat{G}_m[0],\ldots,\hat{G}_m[N-1]]^{\mathsf{T}}$, where $\hat{G}_m[k] = D_m[k]H_m[k]$. We solve the regularized least-squares problem
\begin{align}
    \hat{\mathbf{q}}_m = \arg\min_{\mathbf{q}_m} \bigl\| \mathbf{g}_0 - \mathrm{diag}(\hat{\mathbf{G}}_m)\mathbf{A}\mathbf{q}_m \bigr\|_2^2 + \lambda \|\mathbf{A}\mathbf{q}_m\|_2^2,
    \label{eq:ls_q}
\end{align}
where $\lambda \ge 0$ controls noise amplification, resulting in a closed-form solution $\hat{\mathbf{q}}_m = \mathbf{B}^{-1}\mathbf{A}^{\mathsf{H}} \mathrm{diag}(\hat{\mathbf{G}}_m)^{\mathsf{H}}\mathbf{g}_0$, when $\mathbf{B} = \mathbf{A}^{\mathsf{H}} \mathrm{diag}(\hat{\mathbf{G}}_m)^{\mathsf{H}} \mathrm{diag}(\hat{\mathbf{G}}_m)\mathbf{A} + \lambda \mathbf{A}^{\mathsf{H}}\mathbf{A}$.

\subsection{Timing and Phase Offset Estimation}

We now describe how to estimate the offsets required to construct $D_m[k]$. To estimate $\hat{\tau}_m$ and $\hat{\phi}_m$, we first apply a fractional shift hypothesis $\kappa_m\in[-0.5,0.5]$ to $y_m[n]$. In the frequency domain, this produces
\begin{align}
    Y_{m,\kappa_m}[k] =
    G_m[k]X[k]e^{-j\frac{2\pi k}{N}(\tau_m+\kappa_m)}e^{j\phi_m} + W_m'[k],
   \label{eq:hypothesis_delayed_received_signal}
\end{align}
where $X[k]$, $Y_{m,\kappa_m}[k]$, and $W_m[k]$ are frequency responses of $x[n]$, $y_m[n]$ delayed with $\kappa_m$ samples, and $w_m[n]$, respectively, and $W_m'[k] = e^{-j\frac{2\pi k}{N}\kappa_m}W_m[k]$. We correlate $Y_{m,\kappa_m}[k]$ with the known QPSK pilot, then take the $N$-point IDFT to yield the matched-filter output
\begin{align}
    c_{m,\kappa_m}[n] = \frac{e^{j\phi_m}}{N}
    \sum_{k=0}^{N-1} G_m[k]e^{-j\frac{2\pi k}{N}(\ell_m+\epsilon_m+\kappa_m-n)} + w'_m[n],
    \label{eq:PDP}
\end{align}
where $w'_m[n] =  \sum_{k=0}^{N-1} e^{-j\frac{2\pi k}{N}(\kappa_m-n)}W_m[k]X^*[k] / N$, and $\ell_m\in\mathbb{Z}$ and $\epsilon_m\in[-0.5,0.5]$ are integer and fractional parts of $\tau_m$, respectively, satisfying $\tau_m = \ell_m + \epsilon_m$. Since $G_m[k]$ is strictly real and positive, the magnitude of the noiseless term in \eqref{eq:PDP} is maximized when $\ell_m+\epsilon_m+\kappa_m-n=0$. Accordingly, for each candidate $\kappa_m$, we compute the peak magnitude of $c_{m,\kappa_m}[n]$, and then select the pair $(n, \kappa_m)$ that yields the largest peak. This gives $\hat{\ell}_m=n$ and $\hat{\epsilon}_m=-\kappa_m$, so $\hat{\tau}_m = \hat{\ell}_m+\hat{\epsilon}_m$.

\begin{figure}[t]
    \centering
    \includegraphics[width=0.82\columnwidth]{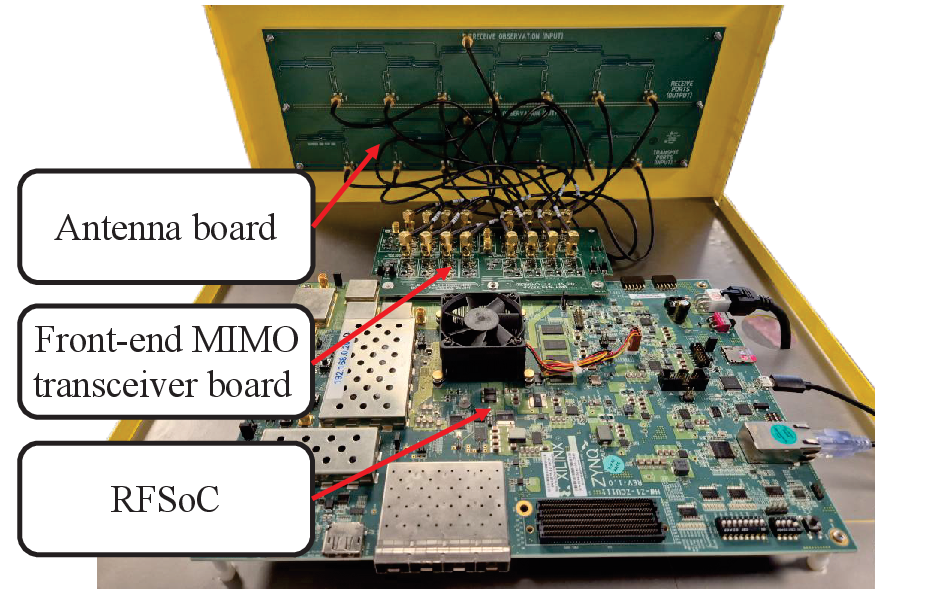}
    \caption{Overall SDR for self-calibration where an RFSoC (bottom) is connected to a front-end MIMO transceiver board (middle), and then on to the antenna board (top).}
    \label{fig:overall structure}
    \vspace{-1.5em}
\end{figure}

If $\hat{\tau}_m=\tau_m$, the matched-filter peak becomes $c_{m,\hat{\epsilon}_m}[\hat{\ell}_m] = e^{j\phi_m}\sum_{k=0}^{N-1}G_m[k] / N + w'_m[\hat{\ell}_m]$. Because $\sum_{k=0}^{N-1}G_m[k]$ is real and positive, the phase of the peak is dominated by $\phi_m$ as long as the noise is negligible. Therefore, the phase-offset estimate is
\begin{align}
    \hat{\phi}_m = \angle c_{m,\hat{\epsilon}_m}[\hat{\ell}_m].
\end{align}
Given $\hat{\tau}_m$, $\hat{\phi}_m$, and $\hat{\mathbf{q}}_m$ obtained from \eqref{eq:ls_q}, we construct the full calibration filter $\{f_m[n]\}_{m=1}^M$ and apply it to the received signal at normal operation.

\section{Hardware description}
\label{sec:hardware description}
The SDR\footnote{Hardware was designed by Pi-Radio and manufactured by Sierra Circuits in Sunnyvale, CA. All hardware and software have been open-sourced at https://github.com/pi-radio/rfsoc-fr1-beamformer} shown in Fig.~\ref{fig:overall structure} is used to realize the self-calibration. It consists of three main components: (i) a Xilinx RFSoC ZCU111 baseband board, (ii) a self-calibration antenna front end, and (iii) a front-end multiple-input multiple-output (MIMO) transceiver board mating the antenna board with the RFSoC.
The goal is to enable fully digital,
squint-free beamforming null forming over the $3.0$-$3.5$~GHz band, a key band for DoD spectrum sharing \cite{NTIA2024SpectrumSharing35GHz}. 
It supports two operating modes: (i) a standalone SDR mode and (ii) an add-on beamforming and frequency-conversion front end for an existing base station.

\begin{figure}[t]
\centering

    \begin{subfigure}{0.47\columnwidth}
    \centering
    \includegraphics[width=\columnwidth]{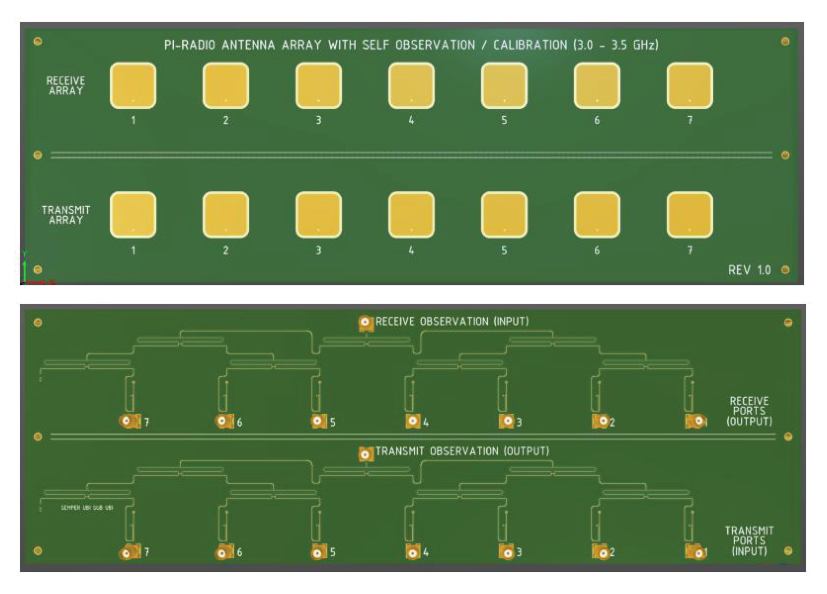}
    \vspace{-1.0em}
    \caption{\centering 3.0-3.5\,\si{GHz} \newline \centering antenna board}
    \label{fig:self-calibration antenna frontend}
    \end{subfigure}
    \vspace{0.1em}
    \begin{subfigure}{0.47\columnwidth}
    \centering
    \includegraphics[width=\columnwidth]{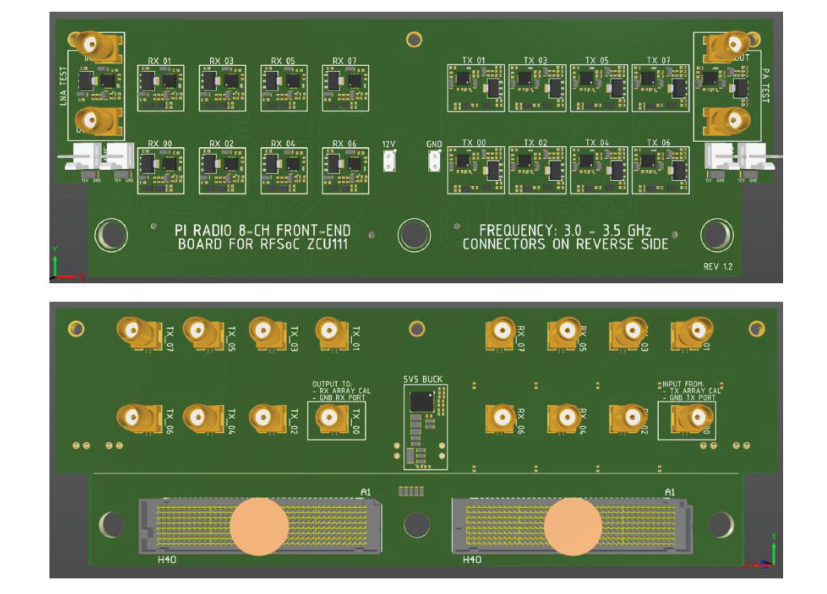}
    \vspace{-1.0em}
    \caption{\centering Front-end MIMO \newline \centering transceiver board}
    \label{fig:SDR_front_back}
    \end{subfigure}
\caption{\textbf{(a)} 
Self-observation antenna board used for self-calibration in the 3.0 to 3.5 \,\si{GHz} bands. The transmission lines for the reference transmitter and receiver are on the \emph{bottom} side of the PCB through a length-matched Wilkinson divider that is directionally coupled to the antenna feeds on the top side.  
\textbf{(b)} The SDR mates to the antenna board through a front-end MIMO transceiver board with standard SMA connectors. }
\vspace{-1.0em} 
\label{fig: antenna frontend and SDR}
\end{figure}

The Xilinx RFSoC ZCU111 board integrates eight 14-bit high-speed digital-to-analog converters and eight 12-bit high-speed analog-to-digital converters (ADCs). It interfaces directly with both the host computer and the front-end MIMO transceiver board. It streams host-generated transmit waveforms to the RF front end and returns captured ADC samples to the host computer for subsequent processing. 
The self-calibration antenna front end, shown in Fig.~\ref{fig:self-calibration antenna frontend}, distributes the reference calibration signal to all channels through a dedicated PCB-based calibration structure, following the structure in Fig.~\ref{fig:rx self calibration}. Rather than using commercial off-the-shelf power combiners and directional couplers, which can introduce non-negligible phase and amplitude imbalance, the proposed front end is fabricated directly on a PCB. This approach reduces mismatch to lithographic tolerances, lowers implementation cost, and provides highly stable signal distribution between the reference node and the calibration nodes. Consequently, the transfer functions across the calibration paths are nearly identical, making the approximation $h'_m[n] \approx h'[n] \text{, }\forall m$ reasonable in practice.  It connects to the RFSoC via the front-end MIMO transceiver board in Fig.~\ref{fig:SDR_front_back}.

The self-calibration procedure for the transmit and receive arrays of the introduced SDR is symmetric. While the receive array calibration we mentioned so far captures all received signals simultaneously when the reference pilot signal is applied, the transmit array calibration must be performed element by element because the reference receive antenna cannot separate simultaneous transmissions from multiple transmit elements without a degradation in the calibration signal-to-interference-plus-noise ratio. Moreover, receive array calibration compensates impairments in post-processing, whereas transmit array calibration applies compensation prior to transmission.

\begin{figure}[t]
\centering
\includegraphics[width=0.85\columnwidth]{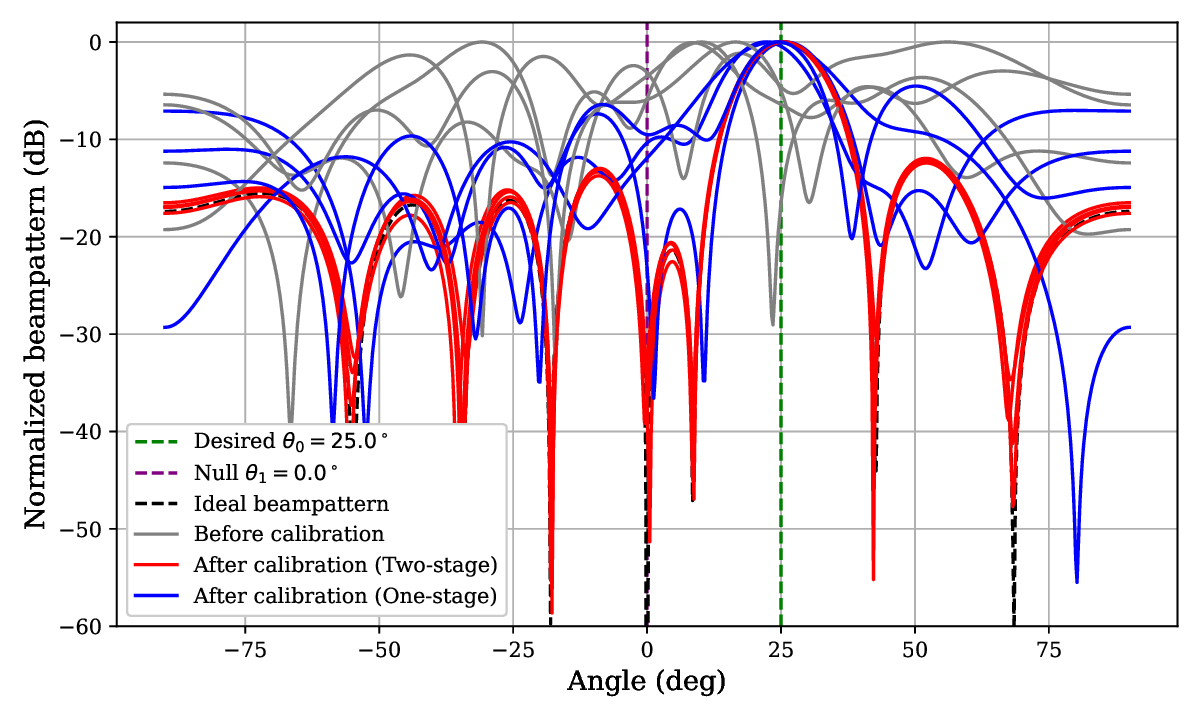}
\vspace{-0.5em}
\caption{\centering Comparison of simulated beampattern pre/post calibration.}
\vspace{-1.7em} 
\label{fig: beampattern pre/post calibration simulation}
\end{figure}

\section{Self-calibration validation}
\label{sec:self-calibration validation}
\subsection{Error Analysis with null forming}

We analyze the performance of the self-calibration in a simple null forming test. In null forming, the receiver aims to maximize the received energy to an angle $\theta_0$ while nulling an angle $\theta_1$. This situation can occur if, for example, the receiver needs to null a strong interferer from $\theta_1$. Ideally, a beamforming vector is
\begin{align}
    \mathbf{b}(\theta_0, \theta_1) = \mathbf{a}(\theta_0) - \frac{\mathbf{a}(\theta_1)^{\mathsf{H}}\mathbf{a}(\theta_0)}{\lVert \mathbf{a}(\theta_1)\rVert^2}\mathbf{a}(\theta_1),
\end{align}
which is a subspace projection result when $\mathbf{a}(\theta) = [a_1(\theta),... , a_{M}(\theta)]^{\mathsf{T}}$. Then, we compute the average nulling ratio $\bar{Q}(\theta_0, \theta_1) = \sum_{k=0}^{N-1} Q(\theta_0, \theta_1, k) / N$. A nulling ratio $Q(\theta_0, \theta_1, k)$ represents the ratio of the channel gain after equalization of the null direction $\theta_1$ relative to the desired direction $\theta_0$ at frequency bin $k$ as
\begin{align}
    Q(\theta_0, \theta_1, k) = \frac{|\mathbf{b}^{\mathsf{H}}(\theta_0, \theta_1)\boldsymbol{\eta}_m(\theta_1, k)|^2}{|\mathbf{b}^{\mathsf{H}}(\theta_0, \theta_1)\boldsymbol{\eta}_m(\theta_0, k)|^2},
\end{align}
where $\boldsymbol{\eta}_m(\theta, k) = F_m[k]H_m[k]\mathbf{a}(\theta)$ is an equalized receiver frequency response.

\begin{figure}[t]
\centering
\includegraphics[width=0.95\columnwidth]{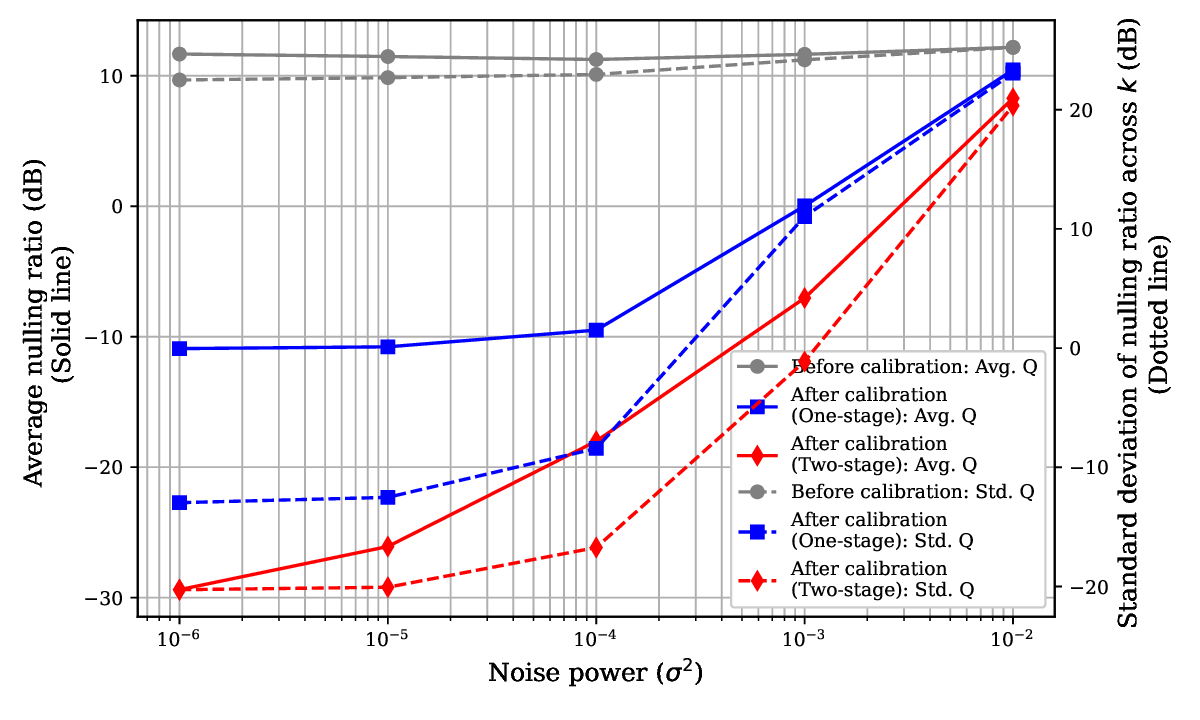}
\vspace{-0.5em}
\caption{\centering Simulated average nulling ratio and standard deviation across different frequency bins w.r.t. noise power.}
\vspace{-1.5em} 
\label{fig: beampattern simulation NMSE}
\end{figure}

\subsection{Simulation Result}
We validate the proposed self-calibration method by comparing the beampatterns before and after calibration in a null forming scenario through computer simulation. In the simulation, we set $N=1024$, $M=8$, $L_d=81$, $L_q=33$, $\lambda=10^{-3}$, $G_0=1$, $\sigma^2=10^{-6}$, and the resolution of $\kappa_m$ to $10^{-2}$, and use a Hamming window for $d_m[n]$. With a beamforming vector designed to steer toward $\theta_0=25^{\circ}$ and place a null at $\theta_1=0^{\circ}$, we plot the beampatterns at $k\in \{422,482,542,602\}$ in Fig.~\ref{fig: beampattern pre/post calibration simulation}. Before calibration, the mainlobes are significantly shifted from $\theta_0$, the nulls at $\theta_1$ are barely formed, and vary across frequency bins, yielding a poor $\bar{Q}(\theta_0,\theta_1)$ of $7.63$~dB. After calibration, the beampatterns become much closer to the ideal null-forming pattern and more consistent across frequency bins, improving $\bar{Q}(\theta_0,\theta_1)$ to $-35.08$~dB. We also compare the proposed filter $f_m[n]$ with an FIR filter of length $L_d+L_q-1$ that directly equalizes \eqref{eq:effective gain}, denoted as one-stage. Its nulling ratio is only $-11.49$~dB, confirming that the proposed two-stage design provides more accurate compensation than the direct one-stage approach.

Fig.~\ref{fig: beampattern simulation NMSE} further quantifies the impact of noise power on the $\bar{Q}(\theta_0, \theta_1)$, for the three cases shown in Fig.~\ref{fig: beampattern pre/post calibration simulation}. The $\bar{Q}(\theta_0, \theta_1)$ degrades as the noise power increases, highlighting the importance of low-noise hardware for accurate null forming. The standard deviation of $Q(\theta_0, \theta_1, k)$ across frequency bins is also shown with dotted lines. After calibration, the frequency selectivity is effectively compensated, reducing the fluctuation of the nulling ratio by about $40$~dB at $\sigma^2=10^{-6}$ compared with the pre-calibration case.

\subsection{Experimental result}
We next evaluate the offset estimation algorithm by comparing the simulated beampatterns before and after calibration based on the measured offsets under the proposed self-calibration SDR architecture using the hardware described in Section~\ref{sec:hardware description}. To isolate the accuracy of the offset estimation algorithm, we compensate for the estimated offsets directly in the frequency domain rather than constructing a time-domain FIR filter. In the experiments, a transmitter with $M=7$ antenna elements sends random QPSK symbols over subcarriers $k \in [412,511] \cup [513,612]$ out of $1024$ total bins, while all remaining bins are set to zero, thus the number of active subcarriers is $N=200$. To reduce over-the-air effects during self-calibration, we ensure that no strong reflectors are present within $5$~m of the SDR. For null forming, we set $\theta_0=25^{\circ}$, $\theta_1=0^{\circ}$, and $G_0=0.1$~dB.

\begin{figure}[t]
\centering
\includegraphics[width=0.85\columnwidth]{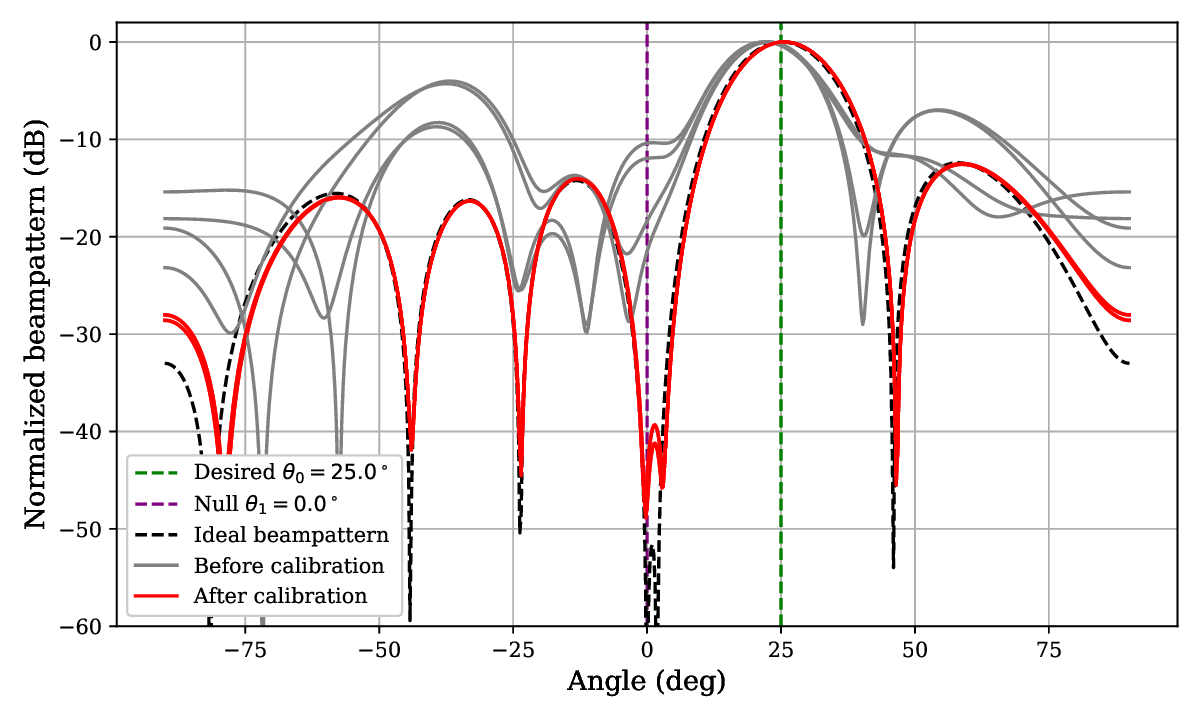}
\vspace{-0.5em}
\caption{\centering Comparison of beampatterns before and after experimental calibration.}
\vspace{-1.5em} 
\label{fig: beampattern pre/post calibration measurement}
\end{figure}

Fig.~\ref{fig: beampattern pre/post calibration measurement} shows the beampatterns before and after the self-calibration procedure at $k \in \{422,482,542,602\}$. Before calibration, the array forms a mainlobe near the desired direction $\theta_0$, but fails to sufficiently suppress the response at $\theta_1$, yielding an average nulling ratio of $\bar{Q}(\theta_0,\theta_1)=-13.12$~dB. In addition, the beampatterns vary across subcarriers, with a standard deviation of $-13.89$~dB, indicating that the formed beams are both inaccurate and unstable. After calibration, the beampatterns exhibit more accurate beam steering, deeper nulling, and improved consistency across frequency bins, achieving an average nulling ratio of $\bar{Q}(\theta_0,\theta_1)=-45.85$~dB with a standard deviation of $-50.33$~dB. This validates the effectiveness of the proposed self-calibration procedure in a practical SDR system.

\section{Conclusion}
\label{sec:conclusion}
This paper presented a self-calibration algorithm based on a physically structured signal model and designed an FIR filter to compensate for per-channel timing, phase, and gain-curve offsets. We validated the algorithm by comparing the null forming performance, which is increasingly essential for spectrum-sharing, anti-jamming, and covert communications in contested and congested environments, but is hard to achieve without calibration, for both the simulation and the experiment. For the experimental validation, we implemented a directionally coupled self-calibration board such that any mismatch across receiver channels originates only from the RFSoC, rather than from the self-calibration board itself. 

\bibliographystyle{IEEEtran}
\bibliography{references}

@string{wsa = "International ITG Workshop on Smart Antennas"}

@article{Dhananjay:2021,
  title={{Pi-Radio v1: Calibration techniques to enable fully-digital beamforming at 60 GHz}},
  author={Dhananjay, Aditya and others},
  journal={Computer Networks},
  volume={196},
  pages={108220},
  year={2021},
  month = {Sep.},
  publisher={Elsevier}
}

@article{kim:2019,
  title={{Fully Digital Beamforming Receiver With a Real-Time Calibration for 5G Mobile Communication}},
  author={Kim, Dong-Chan and others},
  journal={IEEE Transactions on Antennas and Propagation},
  volume={67},
  number={6},
  pages={3809--3819},
  year={2019},
  month = {Mar.},
  publisher={IEEE}
}

@article{Guo:2020,
  title={{A Self-Calibration Method for 5G Full-Digital TDD Beamforming Systems Using an Embedded Transmission Line}},
  author={Guo, Chong and Tian, Ling and Jiang, Zhi Hao and Hong, Wei},
  journal={IEEE Transactions on Antennas and Propagation},
  volume={69},
  number={5},
  pages={2648--2659},
  year={2020},
  month = {Oct.},
  publisher={IEEE}
}

@article{Vaghefi:2020,
  title={{Achieving Phase Coherency and Gain Stability in Active Antenna Arrays for Sub-6 GHz FDD and TDD FD-MIMO: Challenges and Solutions}},
  author={Vaghefi, Reza Monir and others},
  journal={IEEE Access},
  volume={8},
  pages={152680--152696},
  year={2020},
  month = {Aug.},
  publisher={IEEE}
}

@inproceedings{Delfini:2023,
  title={{Impact of the Asymmetric Signal Routing on the Wideband Spatial Behavior of Large Modular Phased Arrays}},
  author={Delfini, Duccio and Tervo, Nuutti and Leinonen, Marko E and P{\"a}rssinen, Aarno},
  booktitle={2023 17th European Conference on Antennas and Propagation (EuCAP)},
  pages={1--5},
  year={2023},
  month = {Mar.},
  organization={IEEE}
}

@article{Tkacenko:2010,
  title={{Wideband Power Amplifier Modeling Incorporating Carrier Frequency Dependent AM/AM and AM/PM Characteristics}},
  author={Tkacenko, Andre},
  journal={Interplanetary Network Progress Report},
  volume={42},
  pages={1--34},
  year={2010}
}

@article{Deng:2020,
  title={{Self-Calibration of Joint RF Impairments in a Loopback Wideband Transceiver}},
  author={Deng, Juinn-Horng and Lee, Chia-Fang and Ku, Meng-Lin and Hwang, Jeng-Kuang},
  journal={IEEE Access},
  volume={8},
  pages={45607--45617},
  year={2020},
  month = {Feb.},
  publisher={IEEE}
}

@inproceedings{Benzin:2017,
  title={{Internal Self-Calibration Methods for Large Scale Array Transceiver Software-Defined Radios}},
  author={Benzin, Andreas and Caire, Giuseppe},
  booktitle={WSA 2017; 21th International ITG Workshop on Smart Antennas},
  pages={1--8},
  year={2017},
  month = {Mar.},
  organization={VDE}
}

@inproceedings{De:2008,
  title={{A Self-Calibrating Quadrature Mixing Front-End for SDR}},
  author={de Witt, Josias J and van Rooyen, Gert-Jan},
  booktitle={2008 IEEE Radio and Wireless Symposium},
  pages={117--120},
  year={2008},
  month = {Jan.},
  organization={IEEE}
}

@inproceedings{kong2020simultaneous,
  title={{Simultaneous Beamforming and Nullforming for Covert Wireless Communications}},
  author={Kong, Justin and Dagefu, Fikadu T and Sadler, Brian M},
  booktitle={VTC2020-Spring},
  year={2020},
  month = {May}
}

@article{jia2025joint,
  title={{Joint Detection, Channel Estimation and Interference Nulling for Terrestrial-Satellite Downlink Co-Existence in the Upper Mid-Band}},
  author={Jia, Shizhen and others},
  journal={arXiv preprint arXiv:2510.08824},
  year={2025},
  month = {Oct.}
}

@article{kumarasiri2025rf,
  title={{RF Anti-Jamming via Multi-Level Howells-Applebaum Null-Forming: 32-Channels, 5.8 GHz/100 MHz/Beam on Xilinx Sx475T FPGA}},
  author={Kumarasiri, Umesha and others},
  journal={IEEE Journal of Radio Frequency Identification},
  year={2025},
  month = {Jun.},
  publisher = {IEEE}
}

@article{tashiro2022nullforming,
  title={{Nullforming-Based Precoder for Spectrum Sharing Between HAPS and Terrestrial Mobile Networks}},
  author={Tashiro, Koji and Hoshino, Kenji and Nagate, Atsushi},
  journal={IEEE Access},
  volume={10},
  pages={55675--55693},
  year={2022},
  month = {May},
  publisher={IEEE}
}

@article{kang2024cellular,
  title={{Cellular Wireless Networks in the Upper Mid-Band}},
  author={Kang, Seongjoon and others},
  journal={IEEE Open Journal of the Communications Society},
  volume={5},
  pages={2058--2075},
  year={2024},
  month = {Mar.},
  publisher={IEEE}
}

@misc{NTIA2024SpectrumSharing35GHz,
  title        = {{NTIA, FCC, Navy Work To Expand Innovative 3.5 GHz Spectrum Sharing Framework}},
  author       = {{National Telecommunications and Information Administration}},
  howpublished = {\url{https://www.ntia.gov/press-release/2024/ntia-fcc-navy-work-expand-innovative-35-ghz-spectrum-sharing-framework}},
  note         = {Press Release},
  year         = {2024},
  month        = {Mar.},
}

@article{Wang:2020,
  title={{Enabling Super-Resolution Parameter Estimation for mm-Wave Channel Sounding}},
  author={Wang, Rui and others},
  journal={IEEE Transactions on Wireless Communications},
  volume={19},
  number={5},
  pages={3077--3090},
  year={2020},
  month = {Feb.},
  publisher={IEEE}
}

@article{Rogalin:2014,
  title={{Scalable Synchronization and Reciprocity Calibration for Distributed Multiuser MIMO}},
  author={Rogalin, Ryan and others},
  journal={IEEE transactions on wireless communications},
  volume={13},
  number={4},
  pages={1815--1831},
  year={2014},
  month = {Mar.},
  publisher={IEEE}
}

@inproceedings{Madani:2021,
  title={{Practical Null Steering in Millimeter Wave Networks}},
  author={Madani, Sohrab and others},
  booktitle={18th USENIX Symposium on Networked Systems Design and Implementation (NSDI 21)},
  pages={903--921},
  year={2021},
  month = {Apr.}
}

@article{Wang:2021,
  title={{Over-the-Air Array Calibration of mmWave Phased Array in Beam-Steering Mode Based on Measured Complex Signals}},
  author={Wang, Zhengpeng and others},
  journal={IEEE Transactions on Antennas and Propagation},
  volume={69},
  number={11},
  pages={7876--7888},
  year={2021},
  month = {May},
  publisher={IEEE}
}

\end{document}